\documentclass{aa} 
\usepackage{graphicx}
\usepackage[varg]{txfonts}
\begin{document}


\title{Which Milky Way masses are consistent with the slightly declining 5-25 kpc rotation curve?}
\titlerunning{Milky Way mass from its rotation curve}

\author{Y. Jiao\inst{1}
\and F. Hammer\inst{1}  
\and J. L. Wang\inst{2}
\and Y. B. Yang\inst{1}
}

\institute{GEPI, Observatoire de Paris, Universit\'e PSL, CNRS, Place Jules Janssen 92195, Meudon, France,  \email{francois.hammer@obspm.fr}
\and 
CAS Key Laboratory of Optical Astronomy, National Astronomical Observatories, Beijing 100101, China}

\date{Received 00 MM 0000 / Accepted 00 MM 0000}


\abstract{Discoveries of extended rotation curves have suggested that spiral galaxy halos contain dark matter. This has led to many studies that estimated the total mass of the Galaxy, mostly using the Navarro, Frenk, and White (NFW) density profile.}{We determine the effect that the choice of the dark matter profile has on the predicted values of extrapolated total masses.}{We considered a recently reported Milky Way (MW) rotation curve, first because of its unprecedented accuracy, and second because the Galactic disk appears to be least affected by past major mergers that have fully reshaped the initial disk.}{We find that the use of an NFW profile (or its generalized form, gNFW) to calculate the dark-matter contribution to the MW rotation curve generates apparently inconsistent results such as an increase in baryonic mass that leads to an increase in dark matter mass.  Furthermore, we find that NFW and gNFW profiles narrow the total mass range, leading to a possible methodological bias particularly against low MW masses. Using the Einasto profile, which is better suited to represent cold dark matter halos, we finally found that the slightly decreasing rotation curve of the MW favors a total mass that can be as low as 2.6 $\times 10^{11}$ $M_{\sun}$, disregarding any other dynamical tracers farther out in the MW.  This is inconsistent with values higher than 18 $\times 10^{11}$ $M_{\sun}$ for any type of cold dark matter halo profiles under the assumption that stars and gas do not affect the predicted dark matter distribution in the MW.}{This methodological paper encourages the use of the Einasto profile to characterize rotation curves with the aim of evaluating their total masses. }



\maketitle



\section{Introduction}
Gaia DR2 provided accurate stellar proper motions to calculate the circular velocity curve of the Milky Way (MW) up to 25 kpc \citep{Eilers2019,Mroz2019}. The result was based on a thorough analysis of a very large sample of 26,000 RGB stars in the MW disk \citep{Eilers2019}, resulting in a slightly but robustly determined decrease in circular velocity from 5 to 25 kpc. While \citet[see also \citealt{Hogg2019}]{Eilers2019} used spectrophotometric distances in their analysis, their finding was confirmed by \cite{Mroz2019} using 773 Classical Cepheids with precise distances. Subsequent analyses of these rotation curves (RCs) have led to a total MW mass near or well below $10^{12} M_{\odot}$ \citep{Eilers2019,deSalas2019,Grand2019,Karukes2020}. \citet{Karukes2020} have used a considerable number of baryonic matter distributions to derive the overall mass distribution, while the \citet{deSalas2019} have accounted for very large error bars after cumulating all the systematics described in details by \citet{Eilers2019}.\\

The accuracy of the MW RC also allows testing different mass profiles for the dark matter (DM) distribution in the MW halo. Recent studies have shown that the three-parameter
Einasto profile \citep[see also \citealt{Retana-Montenegro2012}]{Einasto1965} provides a better description of the CDM halo density profile than the NFW profile \citep{Navarro2004,Navarro2010,Gao2008}, and it is even than the three-parameter generalized gNFW \citep{Klypin2016}.


We propose to test the Einasto and NFW \citep{Navarro1997}) density profiles and their effect on the total mass estimates when spiral rotation curves are fit. We consider the MW RC because of its unprecedented accuracy, and also because the history of the MW is likely quiescent when compared to other spirals \citep{Hammer2007} because the last MW major merger occurred $\sim$10 Gyr ago, as has recently been confirmed based on the resulting debris identified by Gaia DR1 \citep{Belokurov2018} and as will soon be confirmed by Gaia DR2 \citep{Haywood2018,Helmi2018}. 

In Section~\ref{methods} we present our proposed treatment of the error bars for the \citet{Eilers2019} RC, and then describe the choice and mathematical descriptions of the baryon and DM models. In Section~\ref{results} we compare the $\chi^2$ probability distribution for DM represented by the NFW or Einasto profiles. In Section~\ref{discussion} we discuss which mass range is consistent with the combined constraints provided by the fit of the MW RC and by adopting DM halo profiles from the cold dark matter (CDM) theory.


\section{Methods}
\label{methods}
\subsection{Rotation curve and error bars}
\label{sec:errs} 
\citet{Eilers2019} provided  a thorough analysis of the possible systematic errors that may affect the MW RC and summarized (see their Figure 4) four different types of systematics. The first type includes the neglected term in their Jeans equation (see their Equation 3), which is a cross-term made by the vertical density gradient of the product of the radial and vertical velocities. This term is found to be small but not negligible at large distances. For example, \citet{Mackereth2019} showed that vertical velocities are higher for young stars, which is expected because the gaseous disk is likely affected by (former) gas infall. This may affect the derived RC because \cite{Eilers2019} selected relatively young stars ($<$ 4 Gyr) for the MW RC in order to avoid asymmetric drift effects. 

However, the effect is expected to be smaller ($<$ 5 km/s at 12 kpc) than the RC amplitude. The second possible systematics is empirical, and it is an estimate of the error variations with radius after splitting the sample into two parts. We consider here only the first type of systematics because it likely includes the second. 

Adding to this, \citet{Eilers2019} considered a third category of systematics with a quite different nature because it proportionally applies in the same way to all RC points. It is revealed by the three almost horizontal lines in Figure 4 of  \citet{Eilers2019}. This last category of systematics includes the effect of changing the distance of the Sun to the Galactic center, the proper motion of the latter, and it can be extended to the change in  scale length. These uncertainties have to be applied to the derived mass as a whole after the fitting analysis. Added together, they correspond to an additional systematic uncertainty of $\sim$ 2\% on the velocity scale and $\sim$ 4\% on the mass scale. We note in agreement with Christina Eilers (Eilers, 2020, private communication) that summing all the errors of Fig. 4 of \citet{Eilers2019} (as it has been done by \citealt{deSalas2019}) would strongly overestimate the error bars (see the discussion above), which dilutes the significance of the MW RC. 

In the following we adopt the same parameters for the position of the Sun and for the solar velocity as \citet{Eilers2019}. \citet{Karukes2020} have shown that the choice  of the solar velocity may significantly affect the determination of its mass, while it has been considered determined at a 2-3\% level by \citet{Eilers2019}.

\subsection{Milky Way baryonic mass models}
The contribution of the baryonic components to the MW mass or RC is still uncertain, and this may well affect the determination of the DM distribution. Following \citet{Karukes2020}, we adopt here a large number of  models from the literature to describe the MW baryonic component, as described below. The baryonic component and its distribution in the bulge, disk, thick disk, gas, and even halo gas is still debated (see the review by \citealt{Bland-Hawthorn2016}), and some modeling also introduces an ionized gas component \citep{Cautun2020}. The basic idea is to cope with uncertainties on baryons by using a very large grid of possible models,  although we are aware that some baryons models may not be fully consistent with other important constraints from vertical dynamics of the disk stars \citep{Bovy2013} or from microlensing \citep{Wegg2016}.  

\citet{Pouliasis2017} generated a new axisymmetric model (Model I) including a spherical bulge and a thin and thick disk. This model satisfies a number of observational constraints: stellar densities at the solar vicinity, thin- and thick-disk scale lengths and heights, and the absolute value of the perpendicular force $K_z$ as a function of distance to the Galactic center. Although the disk is made of a thin and a thick disk, the associated density profiles are both described by a Miyamoto-Nagai profile (Eq.~\ref{MiNadensityprofile}). 
\citet{Pouliasis2017} concluded that Model I supersedes the axisymmetric model (Model A\&S) proposed by \citet{Allen1991} because there is growing evidence for a strong thick-disk component and because the bulge is less prominent and less classical than assumed in  Model A\&S. Model A\&S consists of a stellar thin disk with a Miyamoto-Nagai profile \citep{Miyamoto1975} and a central bulge with a Plummer profile \citep{Binney2011}. The description of the bulge and disks for both Model I and Model A\&S is expressed in the form \citep{Pouliasis2017} for ($\rm R$, $\rm z$) cylindrical coordinates,
\begin{equation}
\begin{aligned}
\rho_{\operatorname{thin}}(R, z) & = \frac{b_{\operatorname{thin}}^{2} M_{\operatorname{thin}}}{4 \pi}\\
&\times\frac{\left(R^{2} a_{\mathrm{thin}}+3\left(z^{2}+b_{\mathrm{thin}}^{2}\right)^{1 / 2}\right)\left(a_{\mathrm{thin}}+\left(z^{2}+b_{\mathrm{thin}}^{2}\right)^{1 / 2}\right)^{2}}{\left(R^{2}+\left[a_{\mathrm{thin}}+\left(z^{2}+b_{\mathrm{thin}}^{2}\right)^{1 / 2}\right]^{2}\right)^{5 / 2}\left(z^{2}+b_{\mathrm{thin}}^{2}\right)^{3 / 2}}
\label{MiNadensityprofile}
\end{aligned}
\end{equation}
\begin{equation}
\rho_{\mathrm{bulge}}(r)= \frac{3 b_{\mathrm{bulge}}^{2} M_{\mathrm{bulge}}}{4 \pi\left(r^{2}+b_{\mathrm{bulge}}^{2}\right)^{5 / 2}} 
\label{Plummerdensityprofile}
,\end{equation}
where $r=\sqrt{R^2+z^2}$, and $M_{\mathrm{thin}}$, $M_{\mathrm{thick}}$, $M_{\mathrm{bulge}}$, $a_{\mathrm{thin}}$, $a_{\mathrm{thick}}$,  $b_{\mathrm{thin}}$, $b_{\mathrm{thick}}$, $b_{\mathrm{bulge}}$ are the disks and bulge mass and scale constants, respectively (see Table~\ref{table1}).

\begin{figure}
\includegraphics[width=\hsize]{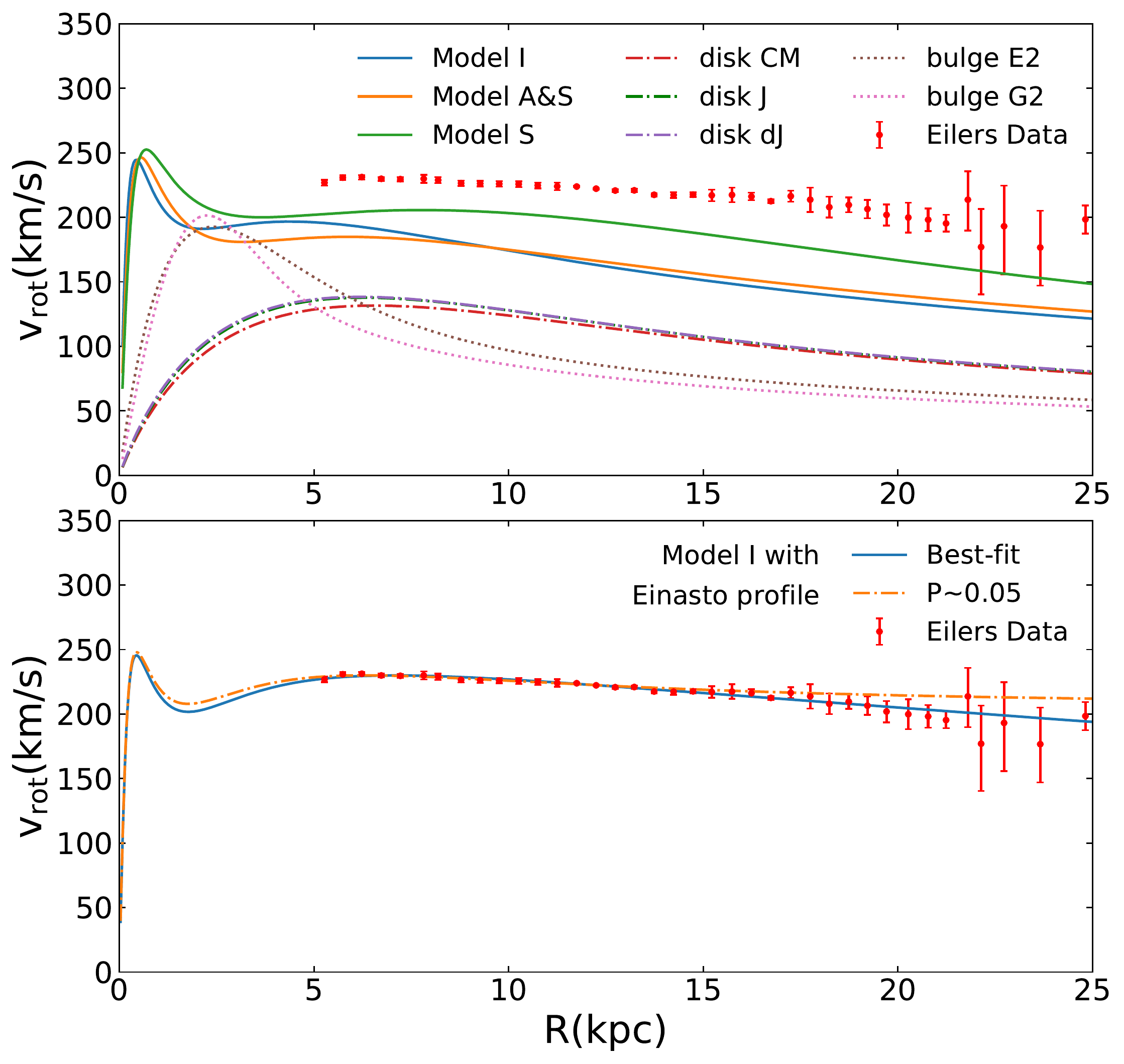}
    \caption{{\it Top: } Contribution to the rotation curve of different baryonic models and model components. Red points indicate the rotation curve of the Milky Way from \citet{Eilers2019}. The error bars are estimated by bootstrapping and include the systematic uncertainties from the neglected term (see text). {\it Bottom: } Fit of the rotation curve by the best-fit model (solid blue curve, total mass of 2.8 $10^{11}$ $M_{\odot}$), and with the most massive MW model for which the $\chi^2$ probability reaches P=0.05 (dash-dotted orange line, total mass of 18 $10^{11}$ $M_{\odot}$), both associated with the baryonic distribution from model I of \citet{Pouliasis2017}.}
    \label{figure1}
\end{figure}

\citet{Sofue2015} presented a model (Model S) of the MW by attempting to fit a `grand rotation curve',which defines the combination of the actual rotation curves (up to 20-25 kpc) with estimates based on orbital motions of objects beyond 25 kpc in the MW halo, e.g., distant globular clusters. The bulge was approximated by a de Vaucouleurs profile \citep{deVaucouleurs1958}. We chose to adopt a Plummer profile (Eq. (\ref{Plummerdensityprofile})) for the bulge, and the disk was assumed to follow an exponentially thin density profile. The surface mass density of the disk is expressed as \citep{Sofue2015}
\begin{equation}
    \Sigma_\mathrm{d}(R)=\Sigma_0\exp{(-R/a_\mathrm{thin})}
    \label{sofuedisk}
,\end{equation}
where $\Sigma_0$ is the central value and $a_\mathrm{thin}$ is the scale radius (see Table~\ref{table1}). This model provides the highest baryonic mass when compared to other models in the literature (see Figure~\ref{figure1}). Nevertheless, we consider it useful for testing the effect of an extremely high baryonic mass for the MW disk and bulge. 

\begin{table}
    \centering
    \caption{Parameters for Model I, Model A\&S, and Model S}
    \begin{tabular}{l l l l }
         \hline\hline
    Parameter & Model I  & Model A\&S & Model S \\
    \hline
    $M_{\mathrm{bulge}} (10^{10} M_{\odot})$ & 1.067 & 1.406 & 2.5 \\
    $M_{\mathrm{thin}} (10^{10} M_{\odot})$ & 3.944 & 8.561 & 11.2 \\
    $M_{\mathrm{thick}} (10^{10} M_{\odot})$ & 3.944 & ---- & ---- \\
    $a_{\mathrm{thin}} (\mathrm{kpc})$ & 5.3 & 5.3178 & 5.73\\
    $a_{\mathrm{thick}} (\mathrm{kpc})$ & 2.6 & ---- & ---- \\
    $b_{\mathrm{bulge}} (\mathrm{kpc})$ & 0.3 & 0.3873 & 0.87\\
    $b_{\mathrm{thin}} (\mathrm{kpc})$ & 0.25 & 0.25 & ---- \\
    $b_{\mathrm{thick}} (\mathrm{kpc})$ & 0.8  &---- &  ----\\
    \hline 
    \end{tabular}
    \label{table1}
\end{table}

A great addition to our choices of baryonic components was presented by \citet{Iocco2015}, and they allowed several possible combinations of models for the bulge and the disk. For the bulge we chose the two triaxial mass density distributions E2 and G2 presented by \citet{Stanek1997}, 
\begin{equation}
    \mathrm{E2} : \rho_{\mathrm{bulge}}(x,y,z)=\rho_0\ e^{-r_1}  
\end{equation}
\begin{equation}
    \mathrm{G2} : \rho_{\mathrm{bulge}}(x,y,z)=\rho_0\ e^{-r_2^2/2} 
,\end{equation}
with \begin{equation}
    r_1^2=\frac{x^2}{x_b^2}+\frac{y^2}{y_b^2}+\frac{z^2}{z_b^2}, r_2^4=\left(\frac{x^2}{x_b^2}+\frac{y^2}{y_b^2}\right)^2+\frac{z^4}{z_b^4}
,\end{equation}
where $(x,y,z)$ are the coordinates along the major, intermediate, and minor axes. For the thin and thick disks, we adopted a double exponential  of the three models (CM from \citealt{Calchi2011}, dJ from \citealt{deJong2010} and J from \citealt{Juric2008}) as described below (see Table~\ref{table3}):
\begin{equation}
\begin{aligned}
    \mathrm{CM} : \rho(R,z)=\Sigma_\mathrm{thin} & \bigg(\frac{1}{2H_1} \exp{\left(-\frac{R}{L_1}-\frac{z}{H_1}\right)}\\ &+f_{\mathrm{thick}}\frac{1}{2H_2}\exp{\left(-\frac{R}{L_2}-\frac{z}{H_2}\right)} \bigg)
\end{aligned}
\end{equation}
\begin{equation}
\begin{aligned}
    \mathrm{J, dJ} : \rho(R,z)=\rho_{\mathrm{thin},\odot} & \bigg( e^{R_\odot/L_1}\exp{\left(-\frac{R}{L_1}-\frac{z}{H_1}\right)}\\ &+f_{\mathrm{thick}}e^{R_\odot/L_2}\exp{\left(-\frac{R}{L_2}-\frac{z}{H_2}\right)} \bigg)
\end{aligned}
.\end{equation}

\begin{table}
\centering
\caption{Parameters for bulge E2 and bulge G2}
\begin{tabular}{l l l}
    \hline\hline
    Parameter & bulge E2 & bulge G2 \\
    \hline
    $M_{\mathbf{bulge}} (10^{10} M_{\odot})$ & 2.41 & 2.12  \\
    $x_b(\mathrm{kpc})$    &0.899&1.239\\
    $y_b(\mathrm{kpc})$    &0.386&0.609\\
    $z_b(\mathrm{kpc})$    &0.250&0.438\\
    \hline
    \end{tabular}
\label{table2}
\end{table}

\begin{table}
\centering
\caption{Parameters of disk CM J and dJ}
\begin{tabular}{l l l l}
     \hline\hline
    Parameter & disk CM  & disk J & disk dJ\\
    \hline
    $M_{\mathbf{thin}} (10^{10} M_{\odot})$ & 3.11&3.17&3.33 \\
    $M_{\mathbf{thick}} (10^{10} M_{\odot})$ &0.82&0.90&0.78\\
    $L_{\mathbf{1}} (\mathrm{kpc})$ &2.75&2.6&2.6\\
    $L_{\mathbf{2}} (\mathrm{kpc})$ &4.1&3.6&4.1\\
    $H_{\mathbf{1}} (\mathrm{kpc})$ &0.25&0.3&0.25\\
    $H_{\mathbf{2}} (\mathrm{kpc})$ &0.75&0.9&0.75\\
    \hline 
    \end{tabular}
\label{table3}
\end{table}

In decreasing order of baryonic mass, Model S, Model A\&S, and then Model I assume significantly higher mass baryonic components than the six combinations of bulge (G2, E2) and disk (CM, dJ, J), which is illustrated by Figure~\ref{figure1} or by comparing Table~\ref{table1} with Tables ~\ref{table2} and ~\ref{table3}.

\subsection{Milky Way dark matter models}
We considered the NFW and Einasto profiles to describe the density profiles of DM halos in spherical coordinates (r). The generalized NFW profile (gNFW, see \citealt{Zhao1996}) can be expressed as in \citet{deSalas2019},
\begin{equation}
    \rho(r)=\frac{\rho_0}{(r/r_0)^{\gamma}(1+r/r_0)^{3-\gamma}}
    \label{NFW}
,\end{equation}
where $\rm r_0$ is the scale radius, and $\rho_0$ is the characteristic dark matter density. For $\gamma$= 1, the profile becomes the NFW profile \citep{Navarro1997} for which we investigate which parameters are able to fit the MW RC, after letting the two NFW parameters, $\rm r_0$ and $\rm m_{NFW}$= 4 $\pi$ $\rho_0$ $\rm r_0^{3}$, vary from 2 to 100 kpc and from 1 to 50 $\times 10^{11}$ $M_{\odot}$, respectively. For the gNFW profile we let the additional parameter, $\gamma$, vary from 0.1 to 3 (see also \citealt{Karukes2020}).
For each tested mass configuration, we verified later that the investigated parameter space was sufficiently large to avoid having missed any solution.

Using the \citet{Retana-Montenegro2012} mathematical framework, the Einasto profile can be written as
\begin{equation}
    \rho(r)=\rho_0\  \mathrm{exp}[-(\frac{r}{h})^{1/n}]
    \label{einasto}
,\end{equation}

where $n$ can determine how fast the density decreases with $r$. To determine which models are able to fit the MW RC, we let the three Einasto parameters, $b_{\rm E}$=3$\times$n, $h_{\rm red}$=$h^{1/n}$, and $\rm m_{E}$= 4 $\pi$ $\rho_0$ $\rm h^{3}$ n $\Gamma(b_{\rm E})$, vary from 3 to 30, 0.05 to 3 and from 1 to 50 $\times 10^{11}$ $M_{\odot}$, respectively. For each tested mass configuration, we verified later that the investigated parameter space was sufficiently large to avoid any missing solution.

In order to determine a non-indefinite total MW mass, the DM halo mass has to be limited by the virial radius, $R_\mathrm{vir}$, which enclosed $M_\mathrm{vir}$, which is the virial mass. We define the virial radius as the radius of the sphere for which the average dark matter density equals 200 times the critical density of the Universe $\rho_\mathrm{cr}$. We adopted a critical density of $\rho_\mathrm{cr}=1.34 \times 10^{-7} M_{\odot}/\mathrm{pc}^3$ , which comes from \citet{Hinshaw2013}. With this definition, the relation between virial radius and virial mass is \begin{equation}
    M_\mathrm{vir}=200\times \frac{4\pi}{3}\rho_\mathrm{cr}R^3_\mathrm{vir}
    \label{virilmass}
.\end{equation}

\section{Results}
\label{results}
\subsection{Deriving the total MW mass and $\chi^2$ probability}
\label{chi2proba}

The total MW potential can be obtained through the Poisson equation,
\begin{equation}
\nabla^2\Phi_\mathrm{tot}=4\pi G\sum_i\rho_\mathrm{i}
,\end{equation}
after adding all the different MW mass components. The theoretical estimate of the circular velocity is derived at different disk radii (R) from the potential $\Phi_\mathrm{tot}$ of the Galaxy through
\begin{equation}
    v_c^2(R)=R\frac{\partial \Phi_\mathrm{tot}}{\partial R}|_{z\approx 0}
    \label{eq1}
.\end{equation}

We applied the $\chi^2$ method to fit the RC and calculate its associated probability, for which we tested an extremely large parameter space. The $\chi^2$ was calculated by the sum at each disk radius $R_i$,
\begin{equation}
    \chi^2=\sum_i^N\frac{(v_{\mathrm{mod},i}-v_{\mathrm{obs},i})^2}{\sigma_i^2}
,\end{equation}
where $v_{\mathrm{mod}}$ is the modeled circular velocity for the cumulative baryons + DM profiles, $v_{\mathrm{obs}}$ is the observed circular velocity, and $\sigma_{stat}$ is the statistical uncertainty of the measurement so that $\sigma_{stat,i}=(\sigma^{+}_{v_{\mathrm{obs},i}}+\sigma^{-}_{v_{\mathrm{obs},i}})/{2,}$ to which we added the systematic uncertainty $\sigma_{sys,i}$ to calculate $\sigma_i$ (see Sect.~\ref{sec:errs} and the table in Appendix A). Hence the $\chi^2$ probability can be expressed as 
\begin{equation}
Prob\left( \frac{ \chi^2}{2}, \frac{N-\nu}{2} \right) = \frac{\gamma\left( \frac{N-\nu}{2}, \frac{ \chi^2}{2} \right)}{\Gamma\left( \frac{N-\nu}{2}\right)}
,\end{equation}
where $N$ is the number of independent observed velocity points in the \citet{Eilers2019} RC, and $\nu$ is the number of degrees of freedom.

\begin{figure}
\includegraphics[width=\hsize]{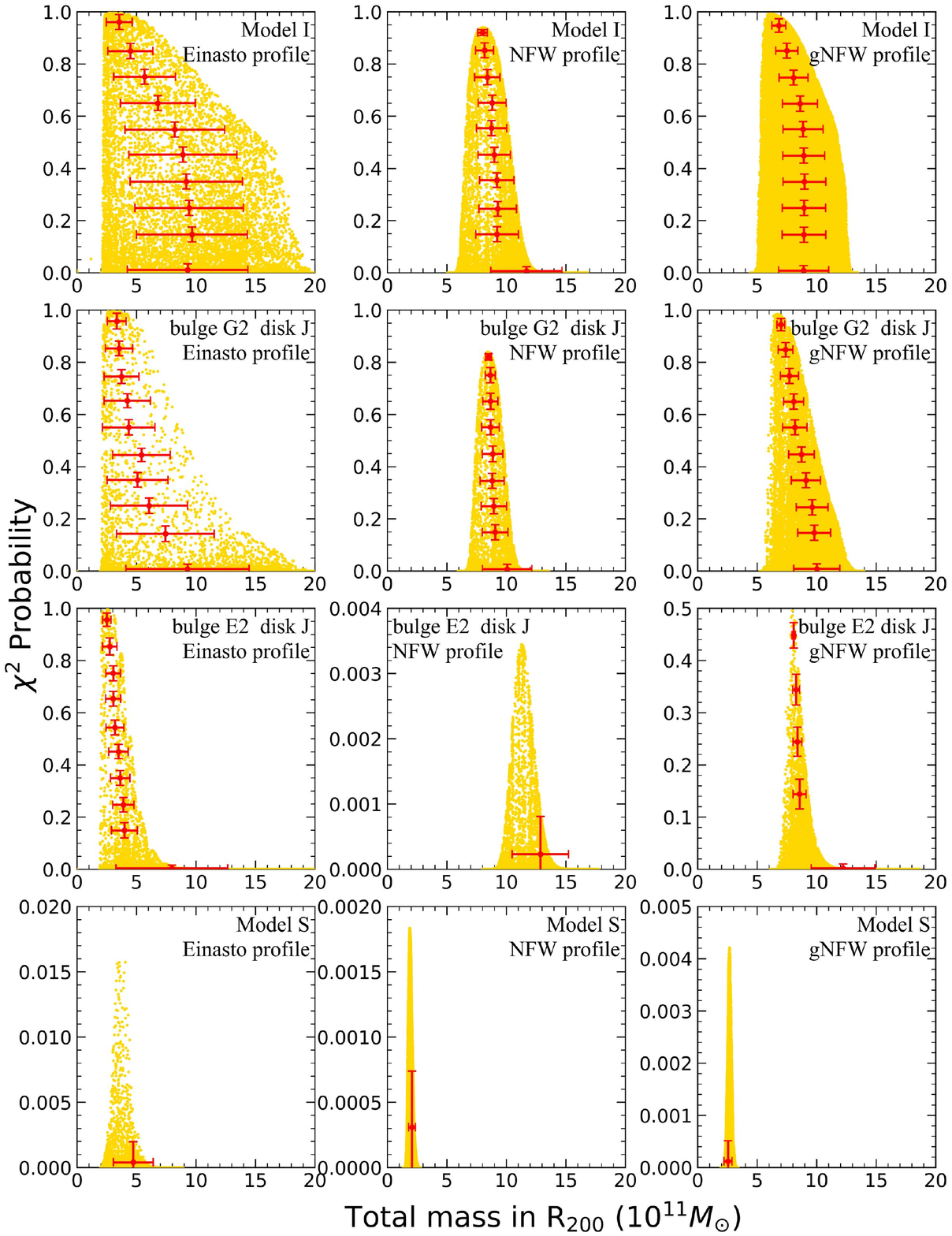}
    \caption{$\chi^2$ probability associated with the combination of different baryonic models with Einasto, NFW, or gNFW mass profiles for the DM. From top to bottom, we show Model I, bulge G2+disk J, bulge E2+disk J, and model S. The red points and error bars are the average and 1$\sigma$ uncertainties.}
    \label{fig:modelIpm}
\end{figure}

To fit the MW RC, we investigated a very large parameter space, allowing for the total MW mass from 1 to 50 $\times$ $10^{11}$ $M_{\odot}$, for instance. In Figure~\ref{fig:modelIpm} each point (P($\chi^2$), $M_{\rm tot}$) represents an investigated baryon+DM model. 
The top panels of Figure~\ref{fig:modelIpm} present the $\chi^2$ probability for the Model I baryon profile \citep{Pouliasis2017} when associated with either the Einasto (left), the NFW (middle), or the gNFW (right) profiles. The first profile shows that high $\chi^2$ probabilities are reached for low MW masses. In contrast, there is no similar trend for the NFW profile, which selects a narrow range of MW masses to fit the RC. The situation is improved with the gNFW, although it does not recover the whole range of masses and especially misses total masses below 5$\times$ $10^{11}$ $M_{\odot}$. The bottom panels present the same for model S, for which the probabilities are very low when associated either with the Einasto or the NFW DM profiles. Examination of the RC fit shows that the baryonic mass is so high that its radial profile is setting up most of the expected RC (see also Figure~\ref{figure1}), leading to differences with the observed RC at almost every radius. This is expected because model S is clearly at odds for the MW; its disk plus bulge mass is higher than that of M31, while half this value is more likely (see, e.g., \citealt{Hammer2007}).

 The two panels in the middle row of Figure~\ref{fig:modelIpm} compare the results when the  bulge is changed from G2 (top) to E2 (bottom), both added to disk J. The first shows a similar behavior as Model I associated with either Einasto or NFW DM profiles. When we used the Einasto profile for the DM, we found that increasing the bulge mass by 15\% (from G2 to E2) is sufficient to exclude high values of the  total mass of the MW. This is expected because when the baryonic mass is increased, a smaller amount of DM mass is available to reproduce the MW RC. Moreover, a too large bulge may limit the number of possible solutions that can fit the RC at low radii. However, for the NFW profile we find that a bulge mass increase from G2 to E2 is sufficient to prevent an efficient reproduction of the MW RC, providing very low $\chi^2$ values. We also find that the associated total (and DM) masses are higher than the mass for the G2 bulge, which disagrees with our expectations. We note that these two properties disappear when the three-parameter gNFW model is used, which might be because for $\gamma < 1$, this profile is less cuspy and is therefore less affected by changes in bulge mass.

The above motivates us to investigate further why adding an additional baryonic mass could lead to an increase of the DM mass when the later is modeled by the NFW density profile. We tested the effect of changing the amount of baryonic mass on the NFW DM mass. We considered a range of baryonic masses scaled on the mass of Model I, with scale factors $\rm f$ varying from 0.85 to 1.15. For $\rm f$= 0.85, 1, and 1.15,  this confirmed that by increasing the baryonic mass, the NFW DM model leads to a significant increase in DM mass from  5.9, 7.2, and 10.7 $\times 10^{11}$ $M_{\odot}$, respectively. This is an unexpected behavior because the DM role is to compensate for the lack of baryonic mass when a given RC is fit. Our first explanation was to relate this to the two-parameter nature of the NFW profile. However, a similar (although less pronounced) behavior affects the gNFW profile. For $\rm f$= 0.85, 1, and 1.15, the gNFW DM model also leads to an increase in DM mass from  4.8, 5.25, and 7.1 $\times 10^{11}$ $M_{\odot}$, respectively. This suggests the following mechanism: for an increasing baryonic mass, the NFW DM scale radius ($r_0$, see Eq.~\ref{NFW}) has to  increase to dilute the DM mass from 5 to 25 kpc (the latest point of the RC). Because outer density slope of the NFW and gNFW is almost constant and shallow (-3) at large radii, this automatically leads to increasing DM masses. This indicates a possible methodological problem of using the NFW (or gNFW) to fit the RC as and to estimate the mass of a galaxy from it.

\subsection{Systematics due to the NFW and gNFW when the total mass is estimated}
To evaluate the differences between Einasto and NFW DM density profiles in fitting the MW RC, we need to ensure that our method does not depend on the initial conditions. In particular, the parameter grid might affect our results because Figure~\ref{fig:modelIpm} shows that the three-parameter space (Einasto or gNFW) might be more difficult to be populated than the two-parameter space (NFW). We further performed for each model a combination of several Monte Carlo simulations that also accounted for the variance due to the RC error bars, which are assumed to follow a Gaussian distribution, in order to fill the high-probability space in the (P($\chi^2$), $M_{\rm tot}$) plane as much as possible.

\begin{figure}
\includegraphics[width=\hsize]{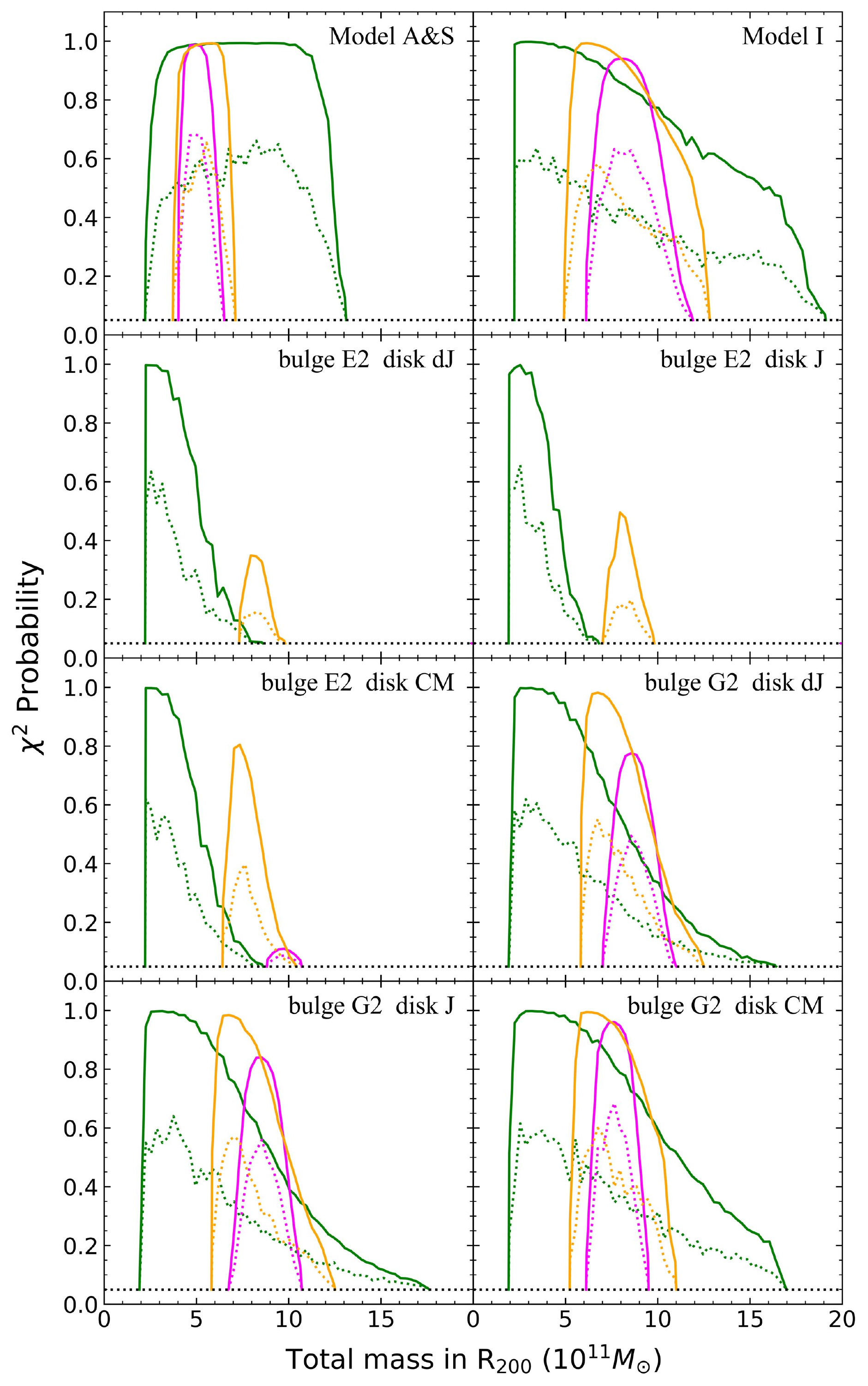}
    \caption{Maximum (solid lines) and averaged (dotted lines) $\chi^2$ probabilities for the different baryonic models. Model names are labeled in each panel, with Einasto, NFW, and gNFW mass predictions in green, magenta, and orange, respectively. The two panels associated with baryonic model E2+dJ and E2+J show no histogram for the NFW because this density profile fails to reproduce the MW RC. The horizontal dotted lines indicates the $\chi^2$ probability limit of 0.05 below which a model is found to be unable to fit the MW RC.}
    \label{fig:proba_tot}
\end{figure}

 The solid lines in Figure~\ref{fig:proba_tot} identify the envelop for each baryonic + DM model, which is defined as being the highest $\chi^2$ probability calculated in mass slices with sizes of 0.3$\times$ $10^{11}$ $M_{\odot}$. We assume that only $\chi^2$ probabilities higher than 0.05 correspond to a good fit of the RC, which we verified after examining the latter. For comparison, Figure~\ref{fig:proba_tot} also shows the averaged probabilities.


Figure~\ref{fig:proba_tot} shows that for all baryonic models, a narrower range of total masses is found for the MW when an NFW or gNFW instead of a Einasto profile is adopted for the DM. Conversely, using the Einasto profile suffices to sample most of the points generated by the NFW profile in the (P($\chi^2$), $M_{\rm tot}$) plane. We find that the total mass solutions based on the NFW  and gNFW profiles are often included in those from the Einasto profile, while using an NFW does not match the highest $\chi^2$ probabilities found by the Einasto model (compare the peaks of the solid magenta and green lines). However, in the case of a massive bulge (E2, especially when associated with disk J), the three-parameter gNFW may sample total MW mass values that cannot be reached by the Einasto model.

Table \ref{tab:estimated_mass} gives the estimated total masses based on the minimum $\chi^2$ values (best fit, highest probabilities) or on averaging the $\chi^2$ probabilities in each mass slice (average). As in Figure~\ref{fig:proba_tot}, the rows are sorted from high- to low-mass baryonic models. This indicates that the best fit of the MW RC for all baryonic models, except for $A\&S$, are unavoidably related to low total masses (from 2.3 to 3.3 $\times 10^{11}$ $M_{\odot}$) if a Einasto profile is chosen for the DM. Conversely, adopting an NFW (or gNFW) profile for the DM leads to much higher total mass values by a factor of 2 to 4. 

 Of the models we studied, Model $A\&S$ possesses the second highest baryonic mass, close to $10^{11}$ $M_{\odot}$, and we investigated why the behavior it shows is so different from that of other baryonic models, especially Model I. In addition to baryonic masses that differ by 11\%, the main difference between the two models is the presence of a thick disk incorporating half the disk mass in Model I, with a scale length that is half that of the thin disk of Models I and $A\&S$. By modifying the thick-disk scale length of Model I to a higher value, we find that this suffices to provide a similar behavior to Model $A\&S$ for the normalized cumulative probabilities of both NFW and Einasto DM profiles. As previously noted for model S, this suggests that an extended and relatively massive baryonic disk determines a significant part of the RC shape.

 Considering the averaged total masses slightly improves the similarities between predictions based on NFW and Einasto DM mass profiles. This is true for Models $A\&S$ and I, which lead to almost consistent NFW and Einasto values of the total masses. However, for lighter baryonic models, the NFW profile for DM still leads to a mass that is higher by factors from 1.5 to 3 when compared to that resulting from the Einasto profile. The NFW (and to a lesser extent, the gNFW) profile appears to preferentially select a narrow range of total masses, excluding in particular the low-mass values that are favored when the Einasto profile is used.

\begin{table*}
        \centering
        \caption{Mass and $\chi^2$ probabilities for Einasto, NFW, and gNFW DM density profiles.}
        \label{tab:estimated_mass}
\begin{tabular}{lllllllllll}
\hline\hline
Baryon  & $M_{\rm bar}$ & $M_{\rm tot}$ & $R_{200}$  &  $M_{\rm tot}$ & $M_{\rm tot}$ & $R_{200}$ & $M_{\rm tot}$ & $M_{\rm tot}$  & $R_{200}$ & $M_{\rm tot}$ \\
model &  &  Einasto & Einasto &  Einasto & NFW & NFW & NFW & gNFW & gNFW & gNFW  \\
 & & best fit & best fit & average & best fit & best fit & average & best fit & best fit & average \\
\hline
S & 1.370 & ---- & ---- & ---- & ---- & ---- & ---- & ---- & ---- & ---- \\
A\&S & 0.997 & $ 7.36^{+3.11}_{-2.85}$ & $186.22^{+20.07}_{-40.21}$ & $ 7.20^{+2.24}_{-2.34}$ & $4.93^{+1.03}_{-0.54}$ & $151,17^{+14.04}_{-12.30}$ & $5.13^{+0.66}_{-0.60}$ & $5.83^{+0.95}_{-0.66}$ & $161.92^{+15.02}_{-12.13}$ & $5.26^{+0.74}_{-0.66}$\\
I & 0.896 & $2.77^{+6.13}_{-0.29}$ & $134.48^{+70.29}_{-9.12}$ & $6.56^{+4.70}_{-3.12}$ & $8.02^{+2.49}_{-0.97}$ & $184.26^{+24.45}_{-13.59}$ & $8.66^{+1.48}_{-1.45}$ & $6.15^{+3.09}_{-0.76}$ & $166.44^{+30.60}_{-13.34}$ & $8.23^{+1.82}_{-1.61}$\\
E2 dJ & 0.652 & $2.37^{+1.03}_{-0.23}$ & $127.63^{+17.75}_{-10.13}$ & $3.01^{+0.70}_{-0.46}$ & ---- & ---- &---- & $7.93^{+0.85}_{-0.70}$ & $186.47^{+8.52}_{-13.45}$ & $8.42^{+0.68}_{-0.65}$\\
E2 J & 0.648 & $2.56^{+1.60}_{-0.13}$ & $130.99^{+17.78}_{-14.46}$ & $3.13^{+0.93}_{-0.82}$ & ---- & ---- & ---- & $7.97^{+1.09}_{-0.74}$ & $186.79^{+9.92}_{-14.62}$ & $8.51^{+0.79}_{-0.75}$\\
E2 CM & 0.634 & $2.41^{+1.30}_{-0.75}$ & $128.34^{+18.66}_{-18.43}$ & $3.59^{+1.01}_{-0.81}$  & $9.68^{+1.45}_{-0.51}$ & $200.24^{+9.77}_{-14.37}$ & $9.72^{+0.80}_{-0.89}$& $7.44^{+1.48}_{-0.61}$ & $182.41^{+15.54}_{-11.74}$ & $8.08^{+0.94}_{-0.93}$\\
G2 dJ & 0.623 & $3.08^{+3.72}_{-0.10}$ & $139.3^{+48.34}_{-7.93}$ & $4.10^{+2.40}_{-1.40}$ & $8.58^{+1.30}_{-1.11}$ & $192.00^{+10.12}_{-18.24}$ & $9.01^{+1.06}_{-1.07}$ & $6.82^{+2.20}_{-0.86}$ & $176.89^{+20.79}_{-14.63}$ & $8.26^{+1.47}_{-1.23}$\\
G2 J & 0.619 & $3.11^{+4.68}_{-0.14}$ & $139.82^{+58.41}_{-6.41}$ & $4.46^{+3.04}_{-1.61}$ & $8.41^{+1.25}_{-1.30}$ & $190.64^{+12.75}_{-16.62}$ & $8.76^{+0.93}_{-1.09}$ & $6.70^{+2.79}_{-0.74}$ & $175.79^{+26.47}_{-13.41}$ & $8.57^{+1.62}_{-1.53}$\\
G2 CM & 0.605 & $3.29^{+4.80}_{-0.28}$ & $142.39^{+59.48}_{-10.11}$ & $5.69^{+3.95}_{-2.49}$ & $7.53^{+1.18}_{-0.93}$ & $183.45^{+12.61}_{-14.05}$ & $7.82^{+0.82}_{-0.89}$& $6.19^{+2.32}_{-0.58}$ & $170.99^{+23.49}_{-11.31}$ & $7.60^{+1.36}_{-1.19}$\\
\hline
\end{tabular}
\tablefoot{Models and associated baryonic mass (first and second columns), and estimated total mass using $\chi^2$ probabilities for Einasto, NFW, and gNFW DM density profiles (third to eighth columns, all masses are given in units of $10^{11} M_{\sun}$). The total mass and mass ranges are evaluated using the minimum $\chi^2$ (best fit, columns 3 and 5) and by weighting the total masses by their $\chi^2$ probabilities (average, columns 4 and 6), together with associated 1$\sigma$ uncertainties. Uncertainties also account for systematics related to the Galactic distance and its motion, as well as to change in scale length ($\sim$ 4\% on masses, see Sect.~\ref{sec:errs}), which have been added to the quoted error bars in this Table. }
\end{table*}

\section{Discussion}
\label{discussion}
\subsection{Limitations of this study and comparison with other works}
The goal of this paper is mostly methodological, that is, we search for the range of total MW masses that  reproduces the MW RC, and then evaluate which mass density profile is the most suitable for estimating the DM mass. We focus on the rotation curve provided by Gaia DR2 alone \citep{Eilers2019,Mroz2019} because its accuracy is several times better than those of any former studies (see Fig. 3 of \citealt{Eilers2019}). This is also because disk stars correspond to dynamical points that are well anchored in the stellar disk, which is assumed to be well in equilibrium with the MW potential. In this context, our study broadens the recent work of \citet{deSalas2019} and \citet{Karukes2020} because here we consider a wider range of baryonic matter models of the MW to fit the Gaia DR2 RC\footnote{The study of \citet{Karukes2020} is not principally based on the Gaia DR2 RC, except   in their Sect. 5.1, in which they favored a similarly low MW mass as in our work (see also their Fig. 8).}. Our resulting total masses for the baryonic Model I are indeed quite similar to the values in Table 2 of \citet{deSalas2019}, thus confirming that using Einasto profile will predict significantly lower total MW masses than when NFW or gNFW profiles are used. Small differences between the two works are probably due to the different schemes in interpreting the systematics of the \citet{Eilers2019} RC. We also retrieved similar results by \citet{Karukes2020}, who also studied the effect of changing the DM density profile. While it goes in the same direction (the Einasto profile predicts lower total masses than the gNFW), their results have not been applied on the accurate Gaia DR2 MW RC, which prevents a detailed comparison.

\begin{figure}
\includegraphics[width=\hsize]{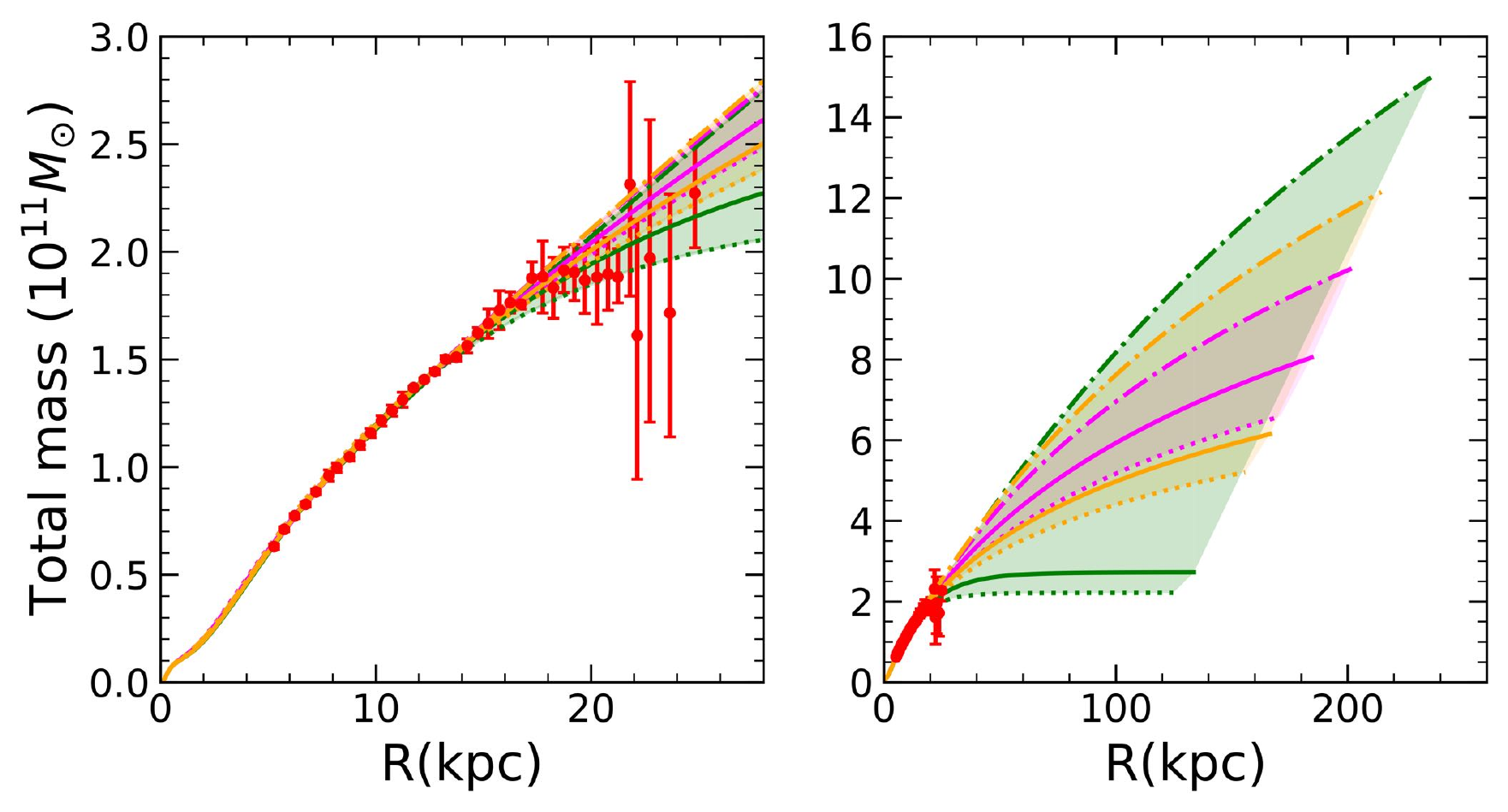}
    \caption{Mass model derived from the MW RC (left panel) and extrapolated to larger radii (right panel), using Model I for the MW baryonic mass.  The best-fit low- and high-mass models are shown as solid, dotted, and dash-dotted lines from Einasto (green), NFW (magenta), and gNFW (orange) models, respectively. Areas showing the possible mass ranges are shaded using the same color code. This shows how the NFW and gNFW bias the mass determination from RCs.}
    \label{fig:mass}
\end{figure}

We are aware that using the RC up to 25 kpc to constrain the mass density profile of the MW is a limited exercise because it needs to be extrapolated to larger radii (see Figure~\ref{fig:mass}). Extrapolations of the total mass from a rotation curve is incorrect, although it has been used very often in the literature either for giant spirals such as the MW (see, e.g., \citealt{Eilers2019} and references therein) of for dwarfs (see, e.g., \citealt{Read2016}). Other works used different mass tracers such as globular clusters \citep{Vasiliev2019}, massive and very bright stars \citep{Deason2020}, or dSph galaxies assumed to be satellites of the MW \citep{Callingham2019}. These methods have the advantage of sampling objects much farther out in the MW halo, although their virial equilibrium with the MW potential is less guaranteed than for rotating disk stars \citep{Eilers2019}. A warp and flare that occur at radii larger than 12 kpc may also limit our study, especially in the outer disk. However, the effect is possibly limited for our $\chi^2$ fitting because the error bars are very large  in the outer disk because they account for the action of the vertical component (see Section~\ref{sec:errs}  and \citealt{Mackereth2019}).

There are two other limitations of our study. The first is linked to the adoption of a spherical halo, although constraints on the dark matter halo shape in the Milky Way are still weak (see \citealt{Read2014}). The second limitation is linked to our choice of initial (flat) priors for DM halo profiles, and this might alter the validity of our results. We compared our initial halos with the \citet{Dutton2014} CDM simulations, in particular, through the relation between concentration and total mass (M200). Our very broad range of parameters encompasses all the \citet[see Figure 3 of \citealt{Udrescu2019}]{Dutton2014} values in the range of $10^{11}-10^{12.5}$ $M_{\odot}$ and in the (c, M200) plane. The solutions that fit the MW rotation curve are also well within the range of halos simulated by \citet{Dutton2014}. 

\begin{figure}
 \includegraphics[width=\hsize]{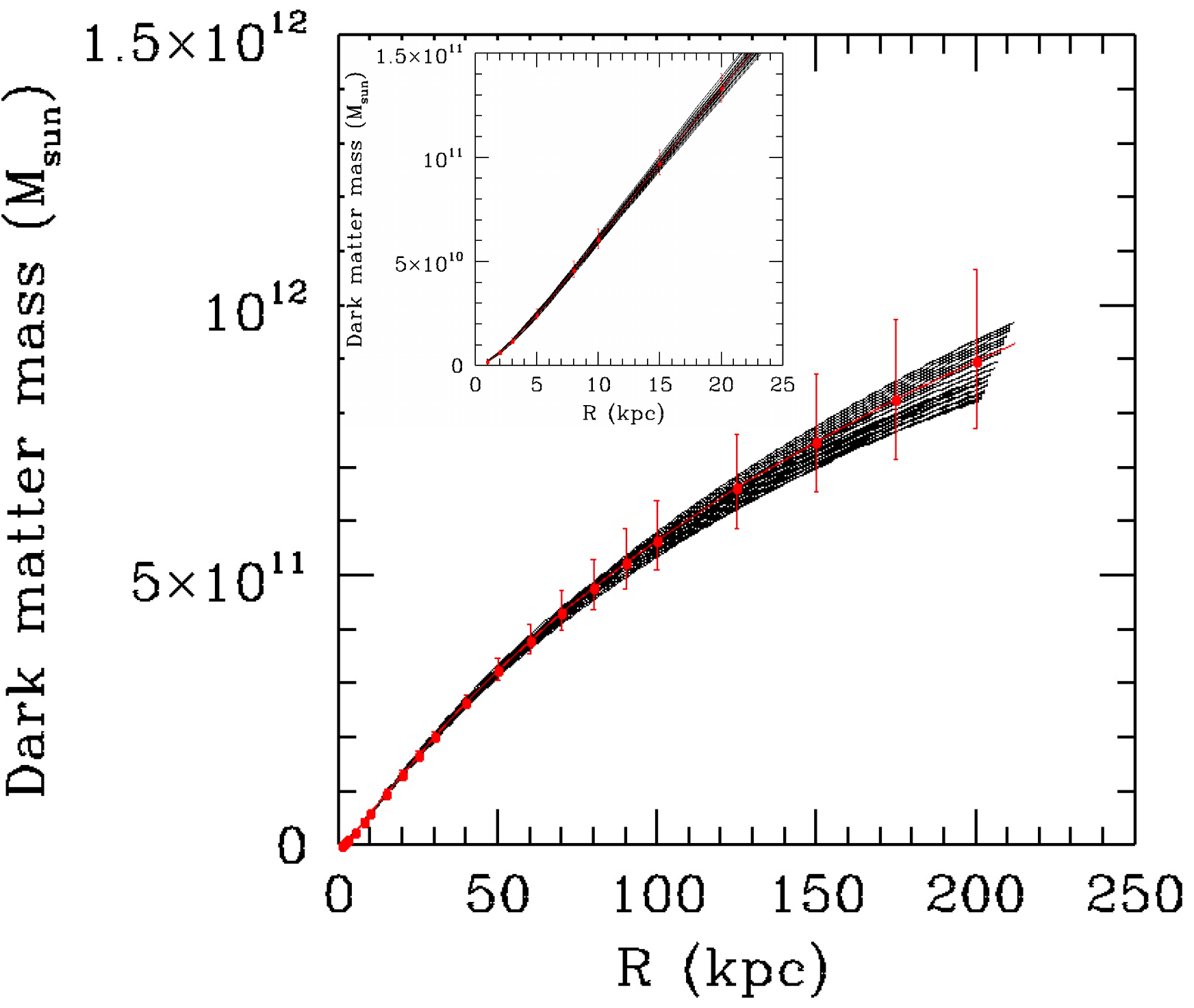}
    \caption{ Dark-matter enclosed mass vs. radius for the mass profile of the contracted halo of \citet{Cautun2020}, from which points and error bars are given in red. When the same baryon content is assumed, the black curves show the result from 24 Einasto models that fit both the RC and the contracted halo. The total mass are very similar within a few percent, and the only small difference is that $R_{200}$ ranges from 200 to 213 kpc instead of 218 kpc for  the contracted halo of \citealt{Cautun2020}. The inset shows a zoom of the mass distribution below 25 kpc to show the similarity of the Einasto DM and the contracted halo near the range of radii of the RC.}
    \label{fig:Cautun_halo_profile}
\end{figure}

Interestingly, our mass boundaries for the $\chi^2$ fitting of the MW RC encompass all these values using other mass tracers. The question remains which mass density profile is the most suitable to properly evaluate the DM contribution to the MW RC. During the submission of this paper, a study by \citet{Cautun2020} was published. They provide a detailed analysis of the effect of baryons on the DM profile. It results in a contracted halo in the spatial region in which the RC is determined. While the total mass is assumed to be consistent within the error bars with the \citet{Callingham2019} value ($M_{\rm tot}$=1.17$\pm$0.18 $\times 10^{12}$ $M_{\odot}$), \citet{Cautun2020} succeeded to fit the MW RC that provided most of the constraints, given its accuracy. Together with our study, this leads us to three important remarks:
 
\begin{itemize}
\item When the MW RC alone is used as a constraint, we find that the Einasto mass density profile leads to the largest range of MW total masses that can reproduce its RC, while both NFW and gNFW profiles lead to a narrow mass range, in particular, by excluding total mass ($M_{200}$=$M_{\rm tot}$) values lower than $\sim$ 5 $\times 10^{11}$ $M_{\odot}$.
\item The contracted halo density profile might be difficult to  reproduce by NFW or by gNFW profiles \citep{Cautun2020}, while it is part of the solutions of this paper using an Einasto profile combined with the baryonic model of \citet{Cautun2020} (see Figure~\ref{fig:Cautun_halo_profile}).
\item We find that both NFW and gNFW profiles provide total masses that increase with baryonic masses (see rows four to nine  in Table~\ref{tab:estimated_mass}), which contradicts expectations that the DM compensates for a lack of mass from baryons in a galaxy. This contradicts the Einasto predictions, according to which the DM mass is higher when the assumed baryonic mass is lower. 
\end{itemize}

It might have been envisioned that these limitations of the NFW profile are related to its two-parameter nature, but this seems to be ruled out by the (almost) similar behavior of the three-parameter gNFW profile. Alternatively, this might be attributed to the density profile of the models in the outskirts. Both have an analytical form that imposes a constant slope of the density profile reaching -3 at large distances, leading to an enclosed mass value that does not converge because it increases as the logarithm of the radius. Investigations by \citet{Nesti2013}  of the internal r$<$ 5kpc MW kinematics showed that a cusp-like NFW (or gNFW for $\gamma$ $>$ 0) profile may also experience some difficulties when combined to baryonic mass.

\subsection{Can the MW has a total mass as low as 2.6 $\times$ 10$^{11}$ $M_{\odot}$?}
\begin{figure}
\includegraphics[width=\hsize]{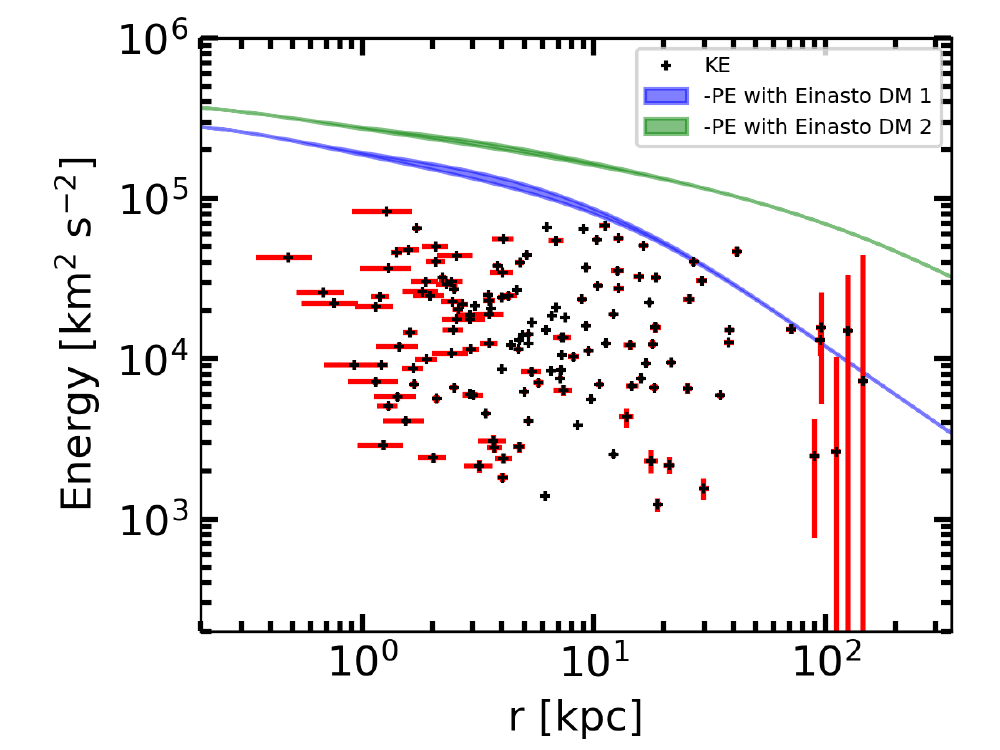}
\caption{Kinetic energy (KE) of the GCs from \citet{Vasiliev2019} (crosses with red error bars) compared to the blue and green thick lines that indicate the potential energy (PE, absolute values) expected from the most likely and the heaviest Einasto model when associated to Model I for baryons, respectively. Error bars have been estimated with Monte Carlo randomly sampling by considering the errors in distance and radial velocity,  as well as errors in proper motion and their covariance. The (small) thickness of the potential lines is due to the presence of the axisymmetric disk component.}
    \label{fig:GCs}
\end{figure}

The Einasto profile fit of the RC points toward low total mass values for the MW (see Table~\ref{tab:estimated_mass}, Figure~\ref{fig:modelIpm} and Figure~\ref{fig:proba_tot}), disregarding any other dynamical tracers farther out in the Milky Way. However, the main result of this paper is provided by the combination of RC fitting with either an Einasto or NFW profile for the MW DM halo, leading to a range of the total MW mass of between 2.5 to 18 $\times 10^{11}$ $M_{\odot}$ (see Figures 1-4). This range is consistent with many studies, including that based on other mass indicators, although they generally disagree with our lowest mass range. 
Figure~\ref{fig:GCs} compares the orbital energy of globular clusters (GCs) from \citet{Vasiliev2019} with that expected from the most likely (total mass: 2.6 $\times 10^{11}$ $M_{\odot}$) and the highest (total mass: 15 $\times 10^{11}$ $M_{\odot}$) MW mass model that could reproduce the MW RC when combining Model I for baryons and  the Einasto profile for DM. Both are consistent with the scenario that GCs are gravitationally bounded to the MW except for one, Pyxis, which appears to disagree significantly for the lighter model. However, the Pyxis eccentric orbit, metallicity, and age indicate an extragalactic origin of Pyxis \citep{Fritz2017}. This indicates that in absence of other precise mass indicators from 25 to 70 kpc, it may be premature to conclude on the total MW mass value from 2.6 $\times 10^{11}$ $M_{\odot}$ and $R_{200}$ = 135 kpc ($\chi^2$) probability =0.999) to 15 $\times 10^{11}$ $M_{\odot}$ and $R_{200}$ = 236 kpc ($\chi^2$ probability =0.35), and even 18 $\times 10^{11}$ $M_{\odot}$ with a $\chi^2$ probability =0.05.

We remark that a low value for the MW mass would have considerable consequences on the orbits of many dSph galaxies, for instance. For example, \citet{Boylan-Kolchin2013} convincingly showed that an MW mass significantly higher than $10^{12}$ $M_{\odot}$ is necessary to bound Leo I.  Using the
\citet{Boylan-Kolchin2013} phase space plot, \citet{Hammer2020} showed that Gaia DR2 orbits might indicate a passage more recent than 4 Gyr ago for many dSphs, assuming a total mass of 8.66 $\times 10^{11}$ $M_{\odot}$ for the MW \citep{Eilers2019}. Because MW dSphs  also have a peculiar planar alignment \citep{Pawlowski2014}, \citet{Deason2020} opted to use halo stars. After a thorough analysis of the possible recent accretions based on phase-space diagrams,  they  derived a total mass within 100 kpc of 6.07 $\times 10^{11}$ $M_{\odot}$\footnote{We do not  discuss their extrapolation to 11.6 $\times 10^{11}$ $M_{\odot}$ for the total MW mass because it depends on the NFW profile that was assumed.}, with which they associated systematics up to 1.2 $\times 10^{11}$ $M_{\odot}$.  This is only marginally consistent with a very low MW mass and makes a future study of Gaia EDR3 results promising that combines the MW RC and GC motions (Wang et al. 2021, in preparation). 

\section{Summary}
Rotation curves are major tools for determining the dynamical mass distribution in the Milky Way and spiral galaxies \citep{Sofue2013}. They are also historically at the root for the requirement of DM in galactic halos \citep{Bosma1978,Rubin1980}, especially when they have been derived from the HI gas, which often extends far beyond the optical disk. Since the end of the 70s, many estimates of the DM content in many spiral galaxies were derived, generally through extrapolations of the observed rotation curves of spiral galaxies.

We have tested the most frequently used density profile to perform numerous analyses of galaxy RCs, namely the NFW density profile \citep{Navarro1997}, and its generalization to three parameters, the gNFW profile. We considered the MW RC because it is one of the most accurately determined RCs \citep{Eilers2019}, and also because the MW has not had a major merger since $\sim$ 10 Gyr \citep{Hammer2007,Helmi2020}. This supports the idea that its disk is dynamically virialized to at least 30 kpc because \citet{Gnedin1999} showed that it takes more than three dynamical times for a system to virialize after a perturbation. 

In contrast to the NFW (or gNFW) profile for DM, the three-parameter Einasto profile \citep[see also \citealt{Retana-Montenegro2012}]{Einasto1965} may account for many types of outer slopes, and it provides a much better fit of the simulated DM properties \citep[and references therein]{Dutton2014}, including for the physically motivated contracted halo \citep{Cautun2020}. It also shows consistent results that can fit the MW RC with most combinations of baryonic mass models, generating a plausible wide range of possible total masses (see Figures  2, 3, and 4). 

 Methodological problems due to the use of a too analytically constrained density model may affect the current estimates of the MW mass such as were reported by \citet{Eilers2019}. Perhaps this also applies to the numerous galaxies for which the RC has been analyzed. Other galaxy RCs have yet to be analyzed using a three-parameter density model for the DM as we did here, although see \cite{Chemin2011} for their promising results. These future investigations should focus first on galaxies that did not experience a recent major merger during which most of the disk was resettled or rebuilt \citep{Hammer2005,Hammer2009,Hopkins2009}. For example, an event like this might complicate the interpretation of the M31 RC, whose recent major merger 2-3 Gyr ago has had a more serious impact (see \citealt{Hammer2018}) than that of the Sagittarius passage near the MW. The Sagittarius passage is thought to have created vertical waves within the MW disk, although this is still disputed (see \citealt{Bennett2020}, and references therein), while the recent merger in M31 has completely destroyed the thin disk of M31 for stars with ages older than 2 Gyr (see the modeling by \citealt{Hammer2018}, which reproduced the anomalous age-velocity dispersion discovered by \citealt{Dorman2015}).  In addition, it is also possible that other two-parameter models are affected in a similar manner, for instance, the isothermal model, which renders comparisons of the validity of these profiles for fitting RCs somewhat obsolete.

Using the Einasto profile, we find that the  MW mass is mostly constrained by its slightly declining RC, which leads to higher $\chi^2$ probabilities for low-mass values (i.e., slightly below 3 $\times 10^{11}$ $M_{\odot}$) for the MW, although less probable higher values up to 18 $\times 10^{11}$ $M_{\odot}$ cannot be excluded. This causes a revision of the available total mass range of the MW down to values that can be as low as 2.6 $\times 10^{11}$ $M_{\odot}$, which are also consistent with the kinetic energy distribution of globular clusters. Further improvements of the accuracy of the MW RC will be invaluable to support or reject these low total masses. They would be invaluable in particular for determining precise orbits for the MW dSphs, for which, given the Gaia EDR3 precision, most uncertainties now come from our insufficient knowledge of the total MW mass.


\begin{acknowledgements}

We are very grateful for the useful and insightful discussions with Christina Eilers about the MW RC and the treatment of the systematics. We warmly thank Piercarlo Bonifacio for his careful advices and comments on the manuscript, and also Frederic Arenou and Carine Babusiaux for their contributions in the meetings during which the methodology of the paper has been adopted.  We are acknowledging the support of the International Research Program Tianguan led by the French CNRS and the Chinese NAOC and Yunnan University.

\end{acknowledgements}



\bibliographystyle{aa}
\bibliography{Jiao_MWmass_V1} 

\begin{thebibliography}{59}
\expandafter\ifx\csname natexlab\endcsname\relax\def\natexlab#1{#1}\fi

\bibitem[{{Allen} \& {Santillan}(1991)}]{Allen1991}
{Allen}, C. \& {Santillan}, A. 1991, \rmxaa, 22, 255

\bibitem[{{Belokurov} {et~al.}(2018){Belokurov}, {Erkal}, {Evans}, {Koposov},
  \& {Deason}}]{Belokurov2018}
{Belokurov}, V., {Erkal}, D., {Evans}, N.~W., {Koposov}, S.~E., \& {Deason},
  A.~J. 2018, \mnras, 478, 611

\bibitem[{{Bennett} \& {Bovy}(2020)}]{Bennett2020}
{Bennett}, M. \& {Bovy}, J. 2020, arXiv e-prints, arXiv:2010.04165

\bibitem[{Binney \& Tremaine(2011)}]{Binney2011}
Binney, J. \& Tremaine, S. 2011, Galactic dynamics (Princeton university press)

\bibitem[{{Bland-Hawthorn} \& {Gerhard}(2016)}]{Bland-Hawthorn2016}
{Bland-Hawthorn}, J. \& {Gerhard}, O. 2016, \araa, 54, 529

\bibitem[{{Bosma}(1978)}]{Bosma1978}
{Bosma}, A. 1978, PhD thesis, Groningen University

\bibitem[{{Bovy} \& {Rix}(2013)}]{Bovy2013}
{Bovy}, J. \& {Rix}, H.-W. 2013, \apj, 779, 115

\bibitem[{{Boylan-Kolchin} {et~al.}(2013){Boylan-Kolchin}, {Bullock}, {Sohn},
  {Besla}, \& {van der Marel}}]{Boylan-Kolchin2013}
{Boylan-Kolchin}, M., {Bullock}, J.~S., {Sohn}, S.~T., {Besla}, G., \& {van der
  Marel}, R.~P. 2013, \apj, 768, 140

\bibitem[{{Calchi Novati} \& {Mancini}(2011)}]{Calchi2011}
{Calchi Novati}, S. \& {Mancini}, L. 2011, \mnras, 416, 1292

\bibitem[{{Callingham} {et~al.}(2019){Callingham}, {Cautun}, {Deason}, {Frenk},
  {Wang}, {G{\'o}mez}, {Grand}, {Marinacci}, \& {Pakmor}}]{Callingham2019}
{Callingham}, T.~M., {Cautun}, M., {Deason}, A.~J., {et~al.} 2019, \mnras, 484,
  5453

\bibitem[{{Cautun} {et~al.}(2020){Cautun}, {Ben{\'\i}tez-Llambay}, {Deason},
  {Frenk}, {Fattahi}, {G{\'o}mez}, {Grand}, {Oman}, {Navarro}, \&
  {Simpson}}]{Cautun2020}
{Cautun}, M., {Ben{\'\i}tez-Llambay}, A., {Deason}, A.~J., {et~al.} 2020,
  \mnras, 494, 4291

\bibitem[{{Chemin} {et~al.}(2011){Chemin}, {de Blok}, \& {Mamon}}]{Chemin2011}
{Chemin}, L., {de Blok}, W.~J.~G., \& {Mamon}, G.~A. 2011, \aj, 142, 109

\bibitem[{{de Jong} {et~al.}(2010){de Jong}, {Yanny}, {Rix}, {Dolphin},
  {Martin}, \& {Beers}}]{deJong2010}
{de Jong}, J. T.~A., {Yanny}, B., {Rix}, H.-W., {et~al.} 2010, \apj, 714, 663

\bibitem[{{de Salas} {et~al.}(2019){de Salas}, {Malhan}, {Freese}, {Hattori},
  \& {Valluri}}]{deSalas2019}
{de Salas}, P.~F., {Malhan}, K., {Freese}, K., {Hattori}, K., \& {Valluri}, M.
  2019, \jcap, 2019, 037

\bibitem[{{de Vaucouleurs}(1958)}]{deVaucouleurs1958}
{de Vaucouleurs}, G. 1958, \apj, 128, 465

\bibitem[{{Deason} {et~al.}(2021){Deason}, {Erkal}, {Belokurov}, {Fattahi},
  {G{\'o}mez}, {Grand}, {Pakmor}, {Xue}, {Liu}, {Yang}, {Zhang}, \&
  {Zhao}}]{Deason2020}
{Deason}, A.~J., {Erkal}, D., {Belokurov}, V., {et~al.} 2021, \mnras, 501, 5964

\bibitem[{{Dorman} {et~al.}(2015){Dorman}, {Guhathakurta}, {Seth}, {Weisz},
  {Bell}, {Dalcanton}, {Gilbert}, {Hamren}, {Lewis}, {Skillman}, {Toloba}, \&
  {Williams}}]{Dorman2015}
{Dorman}, C.~E., {Guhathakurta}, P., {Seth}, A.~C., {et~al.} 2015, \apj, 803,
  24

\bibitem[{{Dutton} \& {Macci{\`o}}(2014)}]{Dutton2014}
{Dutton}, A.~A. \& {Macci{\`o}}, A.~V. 2014, \mnras, 441, 3359

\bibitem[{{Eilers} {et~al.}(2019){Eilers}, {Hogg}, {Rix}, \&
  {Ness}}]{Eilers2019}
{Eilers}, A.-C., {Hogg}, D.~W., {Rix}, H.-W., \& {Ness}, M.~K. 2019, \apj, 871,
  120

\bibitem[{{Einasto}(1965)}]{Einasto1965}
{Einasto}, J. 1965, Trudy Astrofizicheskogo Instituta Alma-Ata, 5, 87

\bibitem[{{Fritz} {et~al.}(2017){Fritz}, {Linden}, {Zivick}, {Kallivayalil},
  {Beaton}, {Bovy}, {Sales}, {Sohn}, {Angell}, {Boylan-Kolchin}, {Carrasco},
  {Damke}, {Davies}, {Majewski}, {Neichel}, \& {van der Marel}}]{Fritz2017}
{Fritz}, T.~K., {Linden}, S.~T., {Zivick}, P., {et~al.} 2017, \apj, 840, 30

\bibitem[{{Gao} {et~al.}(2008){Gao}, {Navarro}, {Cole}, {Frenk}, {White},
  {Springel}, {Jenkins}, \& {Neto}}]{Gao2008}
{Gao}, L., {Navarro}, J.~F., {Cole}, S., {et~al.} 2008, \mnras, 387, 536

\bibitem[{{Gnedin} \& {Ostriker}(1999)}]{Gnedin1999}
{Gnedin}, O.~Y. \& {Ostriker}, J.~P. 1999, \apj, 513, 626

\bibitem[{{Grand} {et~al.}(2019){Grand}, {Deason}, {White}, {Simpson},
  {G{\'o}mez}, {Marinacci}, \& {Pakmor}}]{Grand2019}
{Grand}, R. J.~J., {Deason}, A.~J., {White}, S. D.~M., {et~al.} 2019, MNRAS,
  487, L72

\bibitem[{{Hammer} {et~al.}(2005){Hammer}, {Flores}, {Elbaz}, {Zheng}, {Liang},
  \& {Cesarsky}}]{Hammer2005}
{Hammer}, F., {Flores}, H., {Elbaz}, D., {et~al.} 2005, \aap, 430, 115

\bibitem[{{Hammer} {et~al.}(2009){Hammer}, {Flores}, {Puech}, {Yang},
  {Athanassoula}, {Rodrigues}, \& {Delgado}}]{Hammer2009}
{Hammer}, F., {Flores}, H., {Puech}, M., {et~al.} 2009, \aap, 507, 1313

\bibitem[{{Hammer} {et~al.}(2007){Hammer}, {Puech}, {Chemin}, {Flores}, \&
  {Lehnert}}]{Hammer2007}
{Hammer}, F., {Puech}, M., {Chemin}, L., {Flores}, H., \& {Lehnert}, M.~D.
  2007, \apj, 662, 322

\bibitem[{Hammer {et~al.}(2018)Hammer, Yang, Arenou, Babusiaux, Wang, Puech, \&
  Flores}]{Hammer2018}
Hammer, F., Yang, Y., Arenou, F., {et~al.} 2018, \apj, 860, 76

\bibitem[{{Hammer} {et~al.}(2020){Hammer}, {Yang}, {Arenou}, {Wang}, {Li},
  {Bonifacio}, \& {Babusiaux}}]{Hammer2020}
{Hammer}, F., {Yang}, Y., {Arenou}, F., {et~al.} 2020, \apj, 892, 3

\bibitem[{{Haywood} {et~al.}(2018){Haywood}, {Di Matteo}, {Lehnert}, {Snaith},
  {Khoperskov}, \& {G{\'o}mez}}]{Haywood2018}
{Haywood}, M., {Di Matteo}, P., {Lehnert}, M.~D., {et~al.} 2018, \apj, 863, 113

\bibitem[{{Helmi}(2020)}]{Helmi2020}
{Helmi}, A. 2020, \araa, 58, 205

\bibitem[{{Helmi} {et~al.}(2018){Helmi}, {Babusiaux}, {Koppelman}, {Massari},
  {Veljanoski}, \& {Brown}}]{Helmi2018}
{Helmi}, A., {Babusiaux}, C., {Koppelman}, H.~H., {et~al.} 2018, \nat, 563, 85

\bibitem[{{Hinshaw} {et~al.}(2013){Hinshaw}, {Larson}, {Komatsu}, {Spergel},
  {Bennett}, {Dunkley}, {Nolta}, {Halpern}, {Hill}, {Odegard}, {Page}, {Smith},
  {Weiland}, {Gold}, {Jarosik}, {Kogut}, {Limon}, {Meyer}, {Tucker}, {Wollack},
  \& {Wright}}]{Hinshaw2013}
{Hinshaw}, G., {Larson}, D., {Komatsu}, E., {et~al.} 2013, \apjs, 208, 19

\bibitem[{{Hogg} {et~al.}(2019){Hogg}, {Eilers}, \& {Rix}}]{Hogg2019}
{Hogg}, D.~W., {Eilers}, A.-C., \& {Rix}, H.-W. 2019, \aj, 158, 147

\bibitem[{{Hopkins} {et~al.}(2009){Hopkins}, {Cox}, {Younger}, \&
  {Hernquist}}]{Hopkins2009}
{Hopkins}, P.~F., {Cox}, T.~J., {Younger}, J.~D., \& {Hernquist}, L. 2009,
  \apj, 691, 1168

\bibitem[{{Iocco} {et~al.}(2015){Iocco}, {Pato}, \& {Bertone}}]{Iocco2015}
{Iocco}, F., {Pato}, M., \& {Bertone}, G. 2015, Nature Physics, 11, 245

\bibitem[{{Juri{\'c}} {et~al.}(2008){Juri{\'c}}, {Ivezi{\'c}}, {Brooks},
  {Lupton}, {Schlegel}, {Finkbeiner}, {Padmanabhan}, {Bond}, {Sesar},
  {Rockosi}, {Knapp}, {Gunn}, {Sumi}, {Schneider}, {Barentine}, {Brewington},
  {Brinkmann}, {Fukugita}, {Harvanek}, {Kleinman}, {Krzesinski}, {Long},
  {Neilsen}, {Nitta}, {Snedden}, \& {York}}]{Juric2008}
{Juri{\'c}}, M., {Ivezi{\'c}}, {\v{Z}}., {Brooks}, A., {et~al.} 2008, \apj,
  673, 864

\bibitem[{{Karukes} {et~al.}(2020){Karukes}, {Benito}, {Iocco}, {Trotta}, \&
  {Geringer-Sameth}}]{Karukes2020}
{Karukes}, E.~V., {Benito}, M., {Iocco}, F., {Trotta}, R., \&
  {Geringer-Sameth}, A. 2020, \jcap, 2020, 033

\bibitem[{{Klypin} {et~al.}(2016){Klypin}, {Yepes}, {Gottl{\"o}ber}, {Prada},
  \& {He{\ss}}}]{Klypin2016}
{Klypin}, A., {Yepes}, G., {Gottl{\"o}ber}, S., {Prada}, F., \& {He{\ss}}, S.
  2016, \mnras, 457, 4340

\bibitem[{{Mackereth} {et~al.}(2019){Mackereth}, {Bovy}, {Leung}, {Schiavon},
  {Trick}, {Chaplin}, {Cunha}, {Feuillet}, {Majewski}, {Martig}, {Miglio},
  {Nidever}, {Pinsonneault}, {Aguirre}, {Sobeck}, {Tayar}, \&
  {Zasowski}}]{Mackereth2019}
{Mackereth}, J.~T., {Bovy}, J., {Leung}, H.~W., {et~al.} 2019, \mnras, 489, 176

\bibitem[{{Miyamoto} \& {Nagai}(1975)}]{Miyamoto1975}
{Miyamoto}, M. \& {Nagai}, R. 1975, \pasj, 27, 533

\bibitem[{{Mr{\'o}z} {et~al.}(2019){Mr{\'o}z}, {Udalski}, {Skowron}, {Skowron},
  {Soszy{\'n}ski}, {Pietrukowicz}, {Szyma{\'n}ski}, {Poleski}, {Koz{\l}owski},
  \& {Ulaczyk}}]{Mroz2019}
{Mr{\'o}z}, P., {Udalski}, A., {Skowron}, D.~M., {et~al.} 2019, \apjl, 870, L10

\bibitem[{{Navarro} {et~al.}(1997){Navarro}, {Frenk}, \& {White}}]{Navarro1997}
{Navarro}, J.~F., {Frenk}, C.~S., \& {White}, S. D.~M. 1997, \apj, 490, 493

\bibitem[{{Navarro} {et~al.}(2004){Navarro}, {Hayashi}, {Power}, {Jenkins},
  {Frenk}, {White}, {Springel}, {Stadel}, \& {Quinn}}]{Navarro2004}
{Navarro}, J.~F., {Hayashi}, E., {Power}, C., {et~al.} 2004, \mnras, 349, 1039

\bibitem[{{Navarro} {et~al.}(2010){Navarro}, {Ludlow}, {Springel}, {Wang},
  {Vogelsberger}, {White}, {Jenkins}, {Frenk}, \& {Helmi}}]{Navarro2010}
{Navarro}, J.~F., {Ludlow}, A., {Springel}, V., {et~al.} 2010, \mnras, 402, 21

\bibitem[{{Nesti} \& {Salucci}(2013)}]{Nesti2013}
{Nesti}, F. \& {Salucci}, P. 2013, \jcap, 2013, 016

\bibitem[{{Pawlowski} {et~al.}(2014){Pawlowski}, {Famaey}, {Jerjen}, {Merritt},
  {Kroupa}, {Dabringhausen}, {L{\"u}ghausen}, {Forbes}, {Hensler}, {Hammer},
  {Puech}, {Fouquet}, {Flores}, \& {Yang}}]{Pawlowski2014}
{Pawlowski}, M.~S., {Famaey}, B., {Jerjen}, H., {et~al.} 2014, \mnras, 442,
  2362

\bibitem[{{Pouliasis} {et~al.}(2017){Pouliasis}, {Di Matteo}, \&
  {Haywood}}]{Pouliasis2017}
{Pouliasis}, E., {Di Matteo}, P., \& {Haywood}, M. 2017, \aap, 598, A66

\bibitem[{{Read}(2014)}]{Read2014}
{Read}, J.~I. 2014, Journal of Physics G Nuclear Physics, 41, 063101

\bibitem[{{Read} {et~al.}(2016){Read}, {Iorio}, {Agertz}, \&
  {Fraternali}}]{Read2016}
{Read}, J.~I., {Iorio}, G., {Agertz}, O., \& {Fraternali}, F. 2016, \mnras,
  462, 3628

\bibitem[{{Retana-Montenegro} {et~al.}(2012){Retana-Montenegro}, {van Hese},
  {Gentile}, {Baes}, \& {Frutos-Alfaro}}]{Retana-Montenegro2012}
{Retana-Montenegro}, E., {van Hese}, E., {Gentile}, G., {Baes}, M., \&
  {Frutos-Alfaro}, F. 2012, \aap, 540, A70

\bibitem[{{Rubin} {et~al.}(1980){Rubin}, {Ford}, \& {Thonnard}}]{Rubin1980}
{Rubin}, V.~C., {Ford}, W.~K., J., \& {Thonnard}, N. 1980, \apj, 238, 471

\bibitem[{{Sofue}(2013)}]{Sofue2013}
{Sofue}, Y. 2013, \pasj, 65, 118

\bibitem[{{Sofue}(2015)}]{Sofue2015}
{Sofue}, Y. 2015, \pasj, 67, 75

\bibitem[{{Stanek} {et~al.}(1997){Stanek}, {Udalski}, {Szyma{\'N}ski},
  {Ka{\L}u{\.Z}ny}, {Kubiak}, {Mateo}, \& {Krzemi{\'N}ski}}]{Stanek1997}
{Stanek}, K.~Z., {Udalski}, A., {Szyma{\'N}ski}, M., {et~al.} 1997, \apj, 477,
  163

\bibitem[{{Udrescu} {et~al.}(2019){Udrescu}, {Dutton}, {Macci{\`o}}, \&
  {Buck}}]{Udrescu2019}
{Udrescu}, S.~M., {Dutton}, A.~A., {Macci{\`o}}, A.~V., \& {Buck}, T. 2019,
  \mnras, 482, 5259

\bibitem[{{Vasiliev}(2019)}]{Vasiliev2019}
{Vasiliev}, E. 2019, \mnras, 484, 2832

\bibitem[{{Wegg} {et~al.}(2016){Wegg}, {Gerhard}, \& {Portail}}]{Wegg2016}
{Wegg}, C., {Gerhard}, O., \& {Portail}, M. 2016, \mnras, 463, 557

\bibitem[{{Zhao}(1996)}]{Zhao1996}
{Zhao}, H. 1996, \mnras, 278, 488

\end{thebibliography}


\begin{appendix}

\section{Table A.1 with data of the MW RC and adopted error bars}
Table~\ref{tab:A1}  provides the data for the MW RC given by \citet{Eilers2019}, for which they defined the statistical errors ($\rm \sigma^{-}_{v_c} (km\ s^{-1})$ and $\rm \sigma^{+}_{v_c} (km\ s^{-1})$), and to which we added the systematic error (see Sect.~\ref{sec:errs}) in the last column as a fraction of the observed velocity, following the definition made in Figure 4 of \citet{Eilers2019}.

\begin{table}
        \centering
 \caption{Data points for the MW RC and adopted error bars.}
 \label{tab:A1}
\begin{tabular}{lllll}
\hline\hline
R(kpc) & $\rm v_c (km\ s^{-1})$ & $\rm \sigma^{-}_{v_c} (km\ s^{-1})$ & $\rm \sigma^{+}_{v_c} (km\ s^{-1})$ & $\rm \Delta v_{sys}/v_c) $ \\ 
\hline
5.27 & 226.83 & 1.91 & 1.90 &   0.0045 \\
5.74 & 230.80 & 1.43 & 1.35 &   0.0045 \\
6.23 & 231.20 & 1.70 & 1.10 &   0.0045 \\
6.73 & 229.88 & 1.44 & 1.32 &   0.002 \\
7.22 & 229.61 & 1.37 & 1.11 &   0.0045 \\
7.82 & 229.91 & 0.92 & 0.88 &   0.013 \\
8.19 & 228.86 & 0.80 & 0.67 &   0.010 \\
8.78 & 226.50 & 1.07 & 0.95 &   0.008 \\
9.27 & 226.20 & 0.72 & 0.62 &   0.0088 \\
9.76 & 225.94 & 0.42 & 0.52 &   0.0088 \\
10.26 & 225.68 & 0.44 & 0.40 &  0.010 \\
10.75 & 224.73 & 0.38 & 0.41 &  0.010 \\
11.25 & 224.02 & 0.33 & 0.54 &  0.013 \\
11.75 & 223.86 & 0.40 & 0.39 &  0.001 \\
12.25 & 222.23 & 0.51 & 0.37 &  0.001 \\
12.74 & 220.77 & 0.54 & 0.46 &  0.0046 \\
13.23 & 220.92 & 0.57 & 0.40 &  0.0054 \\
13.74 & 217.47 & 0.64 & 0.51 &  0.0054 \\
14.24 & 217.31 & 0.77 & 0.66 &  0.010 \\
14.74 & 217.60 & 0.65 & 0.68 &  0.0072 \\
15.22 & 217.07 & 1.06 & 0.80 &  0.020 \\
15.74 & 217.38 & 0.84 & 1.07 &  0.0257 \\
16.24 & 216.14 & 1.20 & 1.48 &  0.0123 \\
16.74 & 212.52 & 1.39 & 1.43 &  0.001 \\
17.25 & 216.41 & 1.44 & 1.85 &  0.0182 \\
17.75 & 213.70 & 2.22 & 1.65 &  0.0434 \\
18.24 & 207.89 & 1.76 & 1.88 &  0.0377 \\
18.74 & 209.60 & 2.31 & 2.77 &  0.0247 \\
19.22 & 206.45 & 2.54 & 2.36 &  0.032 \\
19.71 & 201.91 & 2.99 & 2.26 &  0.0385 \\
20.27 & 199.84 & 3.15 & 2.89 &  0.056 \\
20.78 & 198.14 & 3.33 & 3.37 &  0.041 \\
21.24 & 195.30 & 5.99 & 6.50 &  0.010 \\
21.80 & 213.67 & 15.38 & 12.18 & 0.086 \\
22.14 & 176.97 & 28.58 & 18.57 & 0.13 \\
22.73 & 193.11 & 27.64 & 19.05 & 0.13 \\
23.66 & 176.63 & 18.67 & 16.74 & 0.13 \\
24.82 & 198.42 & 6.50 & 6.12 & 0.045 \\
\hline
\end{tabular}
\end{table}
\end{appendix}

\end{document}